\documentclass[aps,prb,twocolumn,showpacs,floatfix]{revtex4}

\usepackage{graphicx}

\begin{document}

\title{The $t$-$J$ model: From light to heavy doping}

\author{A.~Sherman}

\affiliation{Institute of Physics, University of Tartu, Riia 142, 51014
Tartu, Estonia}

\date{\today}

\begin{abstract}
The regions of hole concentrations $0 \leq x \alt 0.3$ and temperatures
$0.005|t| \leq T \leq 0.02|t|$ are studied in the $t$-$J$ model of Cu-O
planes of perovskite high-$T_c$ superconductors. For this purpose
self-energy equations for hole and spin Green's functions are derived
using Mori's projection operator technique and these equations are
self-consistently solved. The calculated hole band transforms radically
at $x\approx 0.08$. A narrow low-concentration band with minima near
$(\pm\frac{\pi}{2},\pm\frac{\pi}{2})$ is converted to a band resembling
the case of weak electron correlations, with the minimum at $(\pi,\pi)$
or $(0,0)$. The hole Fermi surface is respectively changed from small
ellipses at $(\pm\frac{\pi}{2},\pm\frac{\pi}{2})$ to a large rhombus
centered at $(\pi,\pi)$ or $(0,0)$. The decrease of the magnetic
susceptibility at the antiferromagnetic wave vector and spin
correlations with doping is determined by the growth of the frequency
of spin excitations at this momentum. The shape of the frequency
dependence of the susceptibility depends heavily on the hole damping
and varies from a broad feature similar to that observed in
La$_{2-x}$Sr$_x$CuO$_4$ to a pronounced maximum which resembles the
normal-state resonance peak in YBa$_2$Cu$_3$O$_{7-y}$.
\end{abstract}


\maketitle

\section{Introduction}
The two-dimensional $t$-$J$ model was proposed by Anderson
\cite{Anderson} for the description of strong electron correlations in
Cu-O planes of perovskite high-$T_c$ superconductors. In
Ref.~\onlinecite{FZhang} the similarity of the low-energy part of its
spectrum with the spectrum of the realistic three-band Hubbard model
was demonstrated. Nowadays the $t$-$J$ model is one of the most
frequently used models for the interpretation of experimental results
in cuprates (for a review, see Ref.~\onlinecite{Izyumov}). Different
numerical and analytical methods were used for the investigation of the
model. Among these methods are the exact diagonalization of small
clusters, \cite{Dagotto,Bonca} Monte Carlo simulations, \cite{SZhang}
density matrix renormalization group calculations, \cite{White}
spin-wave \cite{Sherman98,Plakida} and mean-field slave-boson
approximations. \cite{Kane} In spite of the considerable progress made
towards the understanding of the properties of the model, the basic
issues of its behavior in going from light to heavy hole doping have
not yet been completely resolved. In particular, there is still no
clear knowledge of how the narrow spin-polaron band inherent in light
doping is converted to a wide band observed in photoemission of
optimally and overdoped crystals. \cite{Damascelli} The investigation
of the variation of the magnetic susceptibility with the hole doping,
temperature and the damping of excitations is also of great importance
in view of a great body of data obtained in inelastic neutron
scattering. \cite{Kastner}

Aiming at a description for the wide range of hole concentrations in
this paper we use the method \cite{Sherman} based on Mori's projection
operator technique. \cite{Mori} The method allows one to derive
self-energy equations for Green's functions constructed from Hubbard
operators without recourse to the intricate diagrammatic technique.
These equations retain the rotation symmetry of spin components
inherent in the Hamiltonian and do not imply any preset magnetic
ordering. In contrast to Ref.~\onlinecite{Sherman}, where this approach
was implemented for the spin subsystem only, in this work we use it
also for the hole subsystem.

The obtained equations were self-consistently solved in a 20$\times$20
lattice for the ranges of hole concentrations $0 \leq x \alt 0.3$ and
temperatures $0.005|t| \leq T \leq 0.02|t|$ where $t$ is the hopping
constant of the $t$-$J$ model. In cuprates this concentration range
covers the regions from light to heavy doping. The ratios of the
exchange $J$ and hopping parameters of the model $J/|t|=0.4$ and
$J/|t|=0.2$ were used. Such ratios correspond to hole-doped cuprates.
\cite{McMahan} For $|t| \approx 0.5$~eV the boundaries of the above
temperature range correspond approximately to 30~K and 120~K.

Obtained results indicate that the hole band transforms radically at
$x\approx 0.08$. A narrow low-concentration band with minima near
$(\pm\frac{\pi}{2},\pm\frac{\pi}{2})$ is converted to a band resembling
in its shape the case of weak electron correlations, with the minimum
at $(\pi,\pi)$ or at $(0,0)$ in dependence on the sign of $t$. The hole
Fermi surface is respectively changed from small elliptical pockets at
$(\pm\frac{\pi}{2},\pm\frac{\pi}{2})$ to a large rhombus centered at
$(\pi,\pi)$ or $(0,0)$. With increasing $x$ the maximum in the
imaginary part of the magnetic susceptibility $\chi''$ at the
antiferromagnetic wave vector ${\bf Q}=(\pi,\pi)$ loses its intensity
and shifts to higher frequencies. This behavior is mainly connected
with the growth of the frequency of spin excitations at {\bf Q} and is
consistent with experimental data. The shape of the frequency
dependence of $\chi''({\bf Q})$ depends heavily on the hole damping and
varies from a broad feature similar to that observed \cite{Aeppli} in
La$_{2-x}$Sr$_x$CuO$_4$ to a pronounced maximum which is analogous to
the normal-state resonance peak \cite{Bourges} in
YBa$_2$Cu$_3$O$_{7-y}$.

\section{Main formulas}
The Hamiltonian of the two-dimensional $t$-$J$ model reads
\begin{equation}\label{hamiltonian}
H=\sum_{\bf nm\sigma}t_{\bf nm}a^\dagger_{\bf n\sigma}a_{\bf
m\sigma}+\frac{1}{2}\sum_{\bf nm}J_{\bf nm}\left(s^z_{\bf n}s^z_{\bf
m}+s^{+1}_{\bf n}s^{-1}_{\bf m}\right),
\end{equation}
where $a_{\bf n\sigma}=|{\bf n}\sigma\rangle\langle{\bf n}0|$ is the
hole annihilation operator, {\bf n} and {\bf m} label sites of the
square lattice, $\sigma=\pm 1$ is the spin projection, $|{\bf
n}\sigma\rangle$ and $|{\bf n}0\rangle$ are site states corresponding
to the absence and presence of a hole on the site. These states may be
considered as linear combinations of the products of the $3d_{x^2-y^2}$
copper and $2p_\sigma$ oxygen orbitals of the extended Hubbard model.
\cite{Jefferson} We take into account nearest neighbor interactions
only, $t_{\bf nm}=t\sum_{\bf a}\delta_{\bf n,m+a}$ and $J_{\bf
nm}=J\sum_{\bf a}\delta_{\bf n,m+a}$ where the four vectors {\bf a}
connect nearest neighbor sites. The spin-$\frac{1}{2}$ operators can be
written as $s^z_{\bf n}=\frac{1}{2}\sum_\sigma\sigma|{\bf
n}\sigma\rangle\langle{\bf n}\sigma|$ and $s^\sigma_{\bf n}=|{\bf
n}\sigma\rangle\langle{\bf n},-\sigma|$.

Properties of the model are determined from the hole and spin retarded
Green's functions
\begin{eqnarray}
&&G({\bf k}t)=-i\theta(t)\langle\!\{a_{\bf k\sigma}(t),a^\dagger_{\bf
k\sigma}\}\!\rangle, \nonumber\\
&&\label{green}\\
&&D({\bf k}t)=-i\theta(t)\langle[s^z_{\bf k}(t),s^z_{\bf -k}]\rangle,
\nonumber
\end{eqnarray}
where $a_{\bf k\sigma}$ and $s^z_{\bf k}$ are the Fourier transforms of
the respective site operators, operator time dependencies and averaging
are defined with the Hamiltonian ${\cal H}=H-\mu\sum_{\bf
n}a^\dagger_{\bf n\sigma}a_{\bf n\sigma}$ with the chemical potential
$\mu$. As mentioned, to obtain self-energy equations for these
functions we used Mori's projection operator technique.
\cite{Mori,Sherman} In this approach the Fourier transform of Green's
function $\langle\!\langle A_0|A^\dagger_0\rangle\!\rangle$ is
represented by the continued fraction
\begin{equation}\label{cfraction}
\langle\!\langle
A_0|A^\dagger_0\rangle\!\rangle=\frac{\displaystyle|A_0\cdot
A_0^\dagger|}{\displaystyle \omega-E_0-\frac{\displaystyle
V_0}{\displaystyle\omega-E_1-\frac{\displaystyle V_1}{\ddots}}}.
\end{equation}
The elements of the fraction $E_i$ and $V_i$ are determined from the
recursive procedure
\begin{eqnarray}
&&[A_n,H]=E_nA_n+A_{n+1}+V_{n-1}A_{n-1},\nonumber\\
&&E_n=|[A_n,H]\cdot A_n^\dagger|\,|A_n\cdot
 A_n^\dagger|^{-1},\nonumber\\
&&\label{lanczos}\\
&&V_{n-1}=|A_n\cdot A_n^\dagger|\,|A_{n-1}\cdot
 A_{n-1}^\dagger|^{-1},\nonumber\\
&&V_{-1}=0,\quad n=0,1,2,\ldots\nonumber
\end{eqnarray}
The operators $A_i$ constructed in the course of this procedure form an
orthogonal set, $|A_i\cdot A^\dagger_j|\propto\delta_{ij}$. In
Eqs.~(\ref{cfraction}) and (\ref{lanczos}) the definition of the inner
product $|A_i\cdot A^\dagger_j|$ depends on the type of the considered
Green's function. For example, for functions (\ref{green}) these are
$\langle\{A_i,A^\dagger_j\}\rangle$ and
$\langle[A_i,A^\dagger_j]\rangle$, respectively. The method described
by Eqs.~(\ref{cfraction}) and (\ref{lanczos}) can be straightforwardly
generalized to the case of many-component operators which is necessary,
for example, to consider Green's functions for Nambu spinors in the
superconducting state. \cite{Sherman}

The residual term of fraction~(\ref{cfraction}) is the Fourier
transform of the quantity
\begin{equation}\label{terminator}
{\cal T}=|A_{nt}\cdot A_n^\dagger|\,|A_{n-1}\cdot
A_{n-1}^\dagger|^{-1},
\end{equation}
where the time evolution of the operator $A_n$ is determined by the
equation
\begin{equation}\label{timevol}
i\frac{d}{dt}{A}_{nt}=\prod_{k=0}^{n-1}(1-P_k)[A_{nt},{\cal H}],\quad
A_{n,t=0}=A_n
\end{equation}
with the projection operators $P_n$ defined as $P_nQ= |Q\cdot
A_n^\dagger|\,|A_n\cdot A_n^\dagger|^{-1}A_n$. The residual term ${\cal
T}$ is a many-particle Green's function which can be estimated by the
decoupling. Following Ref.~\onlinecite{Kondo} the decoupling procedure
may be somewhat improved by introducing the vertex correction $\alpha$
which is determined from the constraint of zero site magnetization
\begin{equation}\label{constraint}
\left\langle s^z_{\bf n}\right\rangle
=\frac{1}{2}\left(1-x\right)-\left\langle s^{-1}_{\bf n}s^{+1}_{\bf
n}\right\rangle=0.
\end{equation}

The spin Green's function obtained in this way reads \cite{Sherman}
\begin{eqnarray}
D({\bf k}\omega)&=&\frac{4(\gamma_{\bf
 k}-1)(JC_1+tF_1)}{\omega^2-\omega\Pi({\bf k}\omega)-
 \omega^2_{\bf k}},\nonumber\\
{\rm Im}\,\Pi({\bf k}\omega)&=&\frac{9\pi t^2J^2(1-x)}{2N(\gamma_{\bf
 k}-1)(JC_1+tF_1)}
 \nonumber\\
&& \label{sgf}\\
&\times&\sum_{\bf k'}(\gamma_{\bf k+k'}-\gamma_{\bf
k'})^2\int^\infty_{-\infty}d\omega' A({\bf k'}\omega')\nonumber\\
&\times&A({\bf k+k'},
 \omega+\omega')\frac{n_F(\omega+\omega')-n_F(\omega')}{\omega},\nonumber
\end{eqnarray}
where $\gamma_{\bf k}=\frac{1}{2}[\cos(k_x)+\cos(k_y)]$, $C_1=\langle
s^{+1}_{\bf n}s^{-1}_{\bf n+a}\rangle$ and $F_1=\langle a^\dagger_{\bf
n}a_{\bf n+a}\rangle$ are the spin and hole correlations on neighboring
sites which can be derived from the respective Green's functions,
\begin{eqnarray}
\langle s^{+1}_{\bf n}s^{-1}_{\bf m}\rangle&=&\frac{2}{N}\sum_{\bf
 k}e^{i{\bf k(n-m)}}\int_{-\infty}^\infty
 d\omega[1+n_B(\omega)]B({\bf k}\omega),\nonumber\\
&&\label{corr}\\
\langle a^\dagger_{\bf n}a_{\bf m}\rangle&=&\frac{1}{N}\sum_{\bf
 k}e^{i{\bf k(n-m)}}\int_{-\infty}^\infty d\omega n_F(\omega)A({\bf
 k}\omega),\nonumber
\end{eqnarray}
\begin{equation}\label{magfreq}
\omega^2_{\bf k}=16J^2\alpha|C_1|(1-\gamma_{\bf k})
(\Delta+1+\gamma_{\bf k}),
\end{equation}
is the square of the frequency of spin excitations with the parameter
$\Delta$ describing the spin gap near {\bf Q}, $N$ is the number of
sites, $A({\bf k}\omega)=-\pi^{-1}{\rm Im}\,G({\bf k}\omega)$ and
$B({\bf k}\omega)=-\pi^{-1}{\rm Im}\,D({\bf k}\omega)$ are the hole and
spin spectral functions, $n_F(\omega)=[\exp(\omega/T)+1]^{-1}$,
$n_B(\omega)=[\exp(\omega/T)-1]^{-1}$ with the temperature $T$. The
real part of the self-energy $\Pi({\bf k}\omega)$ can be calculated
from its imaginary part and the Kramers-Kronig relation. However, in
the calculations ${\rm Re}\,\Pi$ was neglected, since the gap parameter
$\Delta$ and along with it the magnon frequencies near {\bf Q} were
determined from the constraint (\ref{constraint}). This momentum region
plays the main role in the considered spectral and magnetic properties.
As follows from earlier calculations, \cite{Sherman} the vertex
parameter $\alpha$ depends mainly on $x$ and can be approximated as
$\alpha=1.802-0.802\tanh(10x)$.

Two steps of the procedure (\ref{cfraction}), (\ref{lanczos}) can be
also carried out for the hole Green's function. We find
\begin{eqnarray}
&&\langle\{a_{\bf k\sigma},a^\dagger_{\bf k\sigma}\}\rangle=
 \frac{1}{2}(1+x)=\phi,\nonumber\\
&&E_0=(4t\phi+6tC_1\phi^{-1}-3JF_1\phi^{-1})\gamma_{\bf k}
 \nonumber\\
&&\mbox{ }+4tF_1\phi^{-1}-3JC_1\phi^{-1}-\mu,\nonumber\\
&&V_0=24t^2C'_{\bf k}+16t^2(3C_1+\phi^2)\gamma_{\bf k} \nonumber\\
&&\label{hcf} \\
&&\mbox{ }+t^2\left[\frac{1}{4}x(1-x)-4C_1\left(1+\frac{3}{2}x\right)
 \phi^{-1}\right]\nonumber\\
&&\mbox{ }+24t^2F_1\gamma_{\bf k}-4t(3C_1\phi^{-1}+2\phi)(E_0+\mu)
 \gamma_{\bf k}\nonumber\\
&&\mbox{ }-8tF_1(E_0+\mu)+(E_0+\mu)^2,\nonumber\\
&&E_1\approx -\mu, \nonumber
\end{eqnarray}
where $x=N^{-1}\sum_{\bf k}\int_{-\infty}^\infty d\omega
n_F(\omega)A({\bf k}\omega)$ and $C'_{\bf k}=2N^{-1}$ $\sum_{\bf
k'}\gamma^2_{\bf k-k'}\langle s^z_{\bf k'}s^z_{\bf -k'}\rangle$. If the
continued fraction (\ref{cfraction}) is terminated at this stage, the
hole spectrum is approximated by two poles which form two bands,
\begin{eqnarray}
G({\bf
k}\omega)&=&\frac{\phi(\mu-\omega)}{(\widetilde{E}^2_0+V_0)^{1/2}}
\left(\frac{1}{\omega-\varepsilon_{{\bf k},1}+\mu+i\eta}\right.\nonumber\\
&-&\left.\frac{1}{\omega-\varepsilon_{{\bf
k},2}+\mu+i\eta}\right),\label{hgf}
\end{eqnarray}
where
\begin{eqnarray}
&&\widetilde{E}_0=(4t\phi+6tC_1\phi^{-1}-3JF_1\phi^{-1})\gamma_{\bf
k},\nonumber\\
&&\label{energy}\\
&&\varepsilon_{{\bf
k},j}=\widetilde{E}_0/2\pm(\widetilde{E}_0^2/4+V_0)^{1/2},\nonumber
\end{eqnarray}
and $\eta$ is a damping of the hole states. In the framework of the
$t$-$J$ model this damping is connected with the hole-magnon
scattering. \cite{Sherman} However, in this work the damping is
considered as a free parameter to take into account other possible
damping processes. Such approach is motivated by the fact that the
damping of spin excitations (\ref{sgf}) depends heavily on the value of
$\eta$ and to investigate different shapes of the frequency dependence
of the susceptibility $\eta$ is considered as a variable parameter.

Thus, in the present work the calculation of the elements of the
continued fraction $E_i$ and $V_i$ for the spin Green's function was
carried out up to the second order: $E_0$, $V_0$ and $E_1$ were
calculated. Then the residual term of the continued fraction (the part
of the fraction containing the third and higher order terms) which is
represented by the many-particle Green's function (\ref{terminator}) is
approximated by the decoupling. The same elements of the continued
fraction were calculated for the hole Green's function [see
Eq.~(\ref{hcf})]. In this case the residual part of the continued
fraction was omitted and the artificial damping $\eta$ was added. The
main argument for the above truncations of the continued fractions is
based on the results obtained previously. In Ref.~\onlinecite{Kondo}
the analogous truncation in the equations of motion with the decoupling
improved by a vertex correction was successfully used for the
description of spin excitations in the Heisenberg model. For the
lightly doped $t$-$J$ model this approximation was used in
Ref.~\onlinecite{Sherman} and the results obtained were in good
agreement with the exact diagonalization of small clusters. The
two-pole approximation for the hole Green's function was shown to give
a good description of the spin-polaron band for moderate $|t|/J$ in the
spin-wave approach. \cite{Sherman98} In a more general perspective the
used method is a version of the Lanczos algorithm. It is known from
numerical methods that at a successful choice of the starting vector
this algorithm gives good approximations for outermost eigenvalues and
eigenvectors of a matrix already with few steps.

Equations (\ref{constraint})--(\ref{hgf}) form a closed set which can
be solved by iteration for given values of the chemical potential,
temperature, hole damping and the sign of $t$.

\section{Evolution of the hole spectrum}
\begin{figure}
\includegraphics[width=6.5cm]{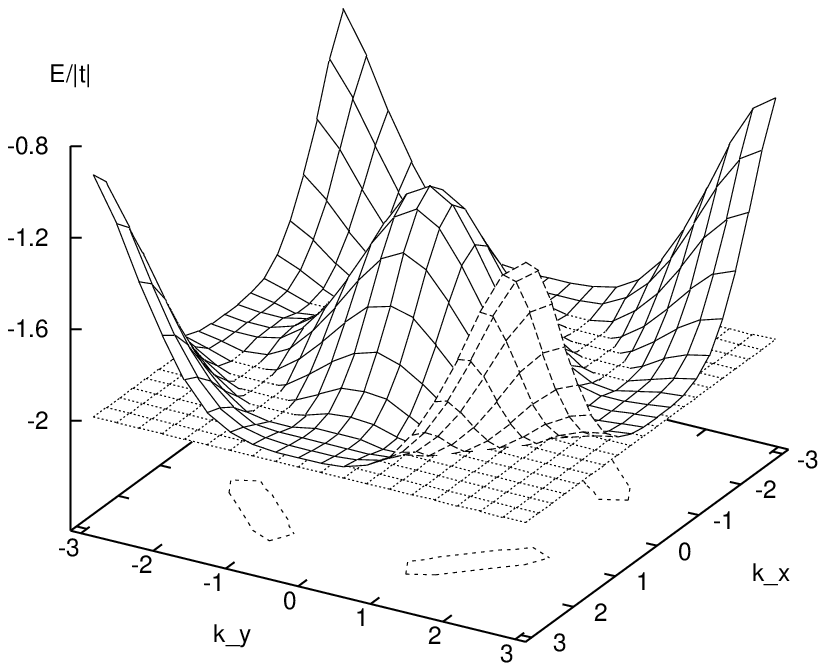}
\includegraphics[width=6.5cm]{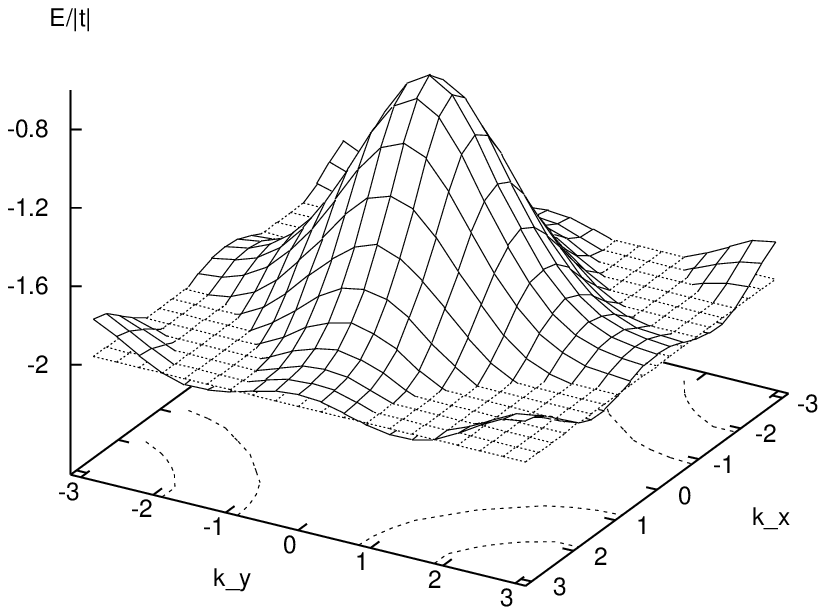}
\includegraphics[width=6.5cm]{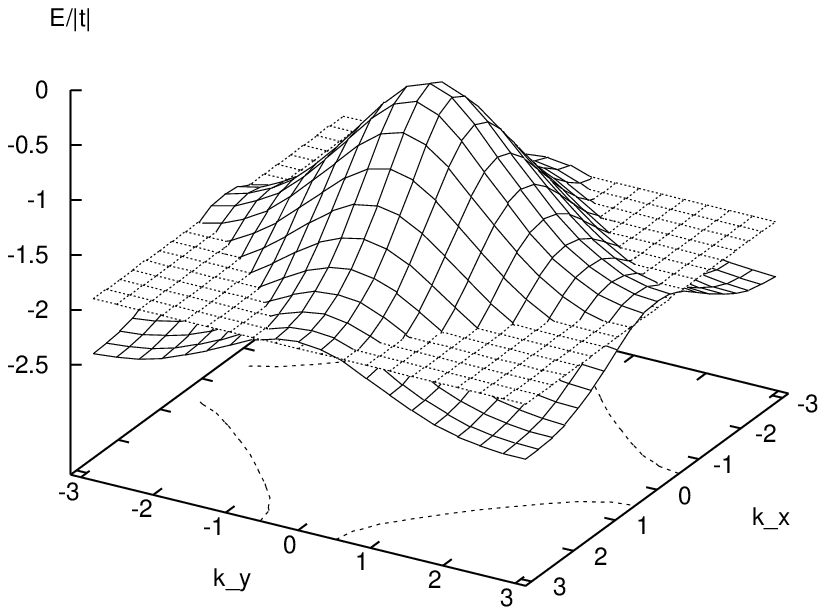}
\includegraphics[width=6.5cm]{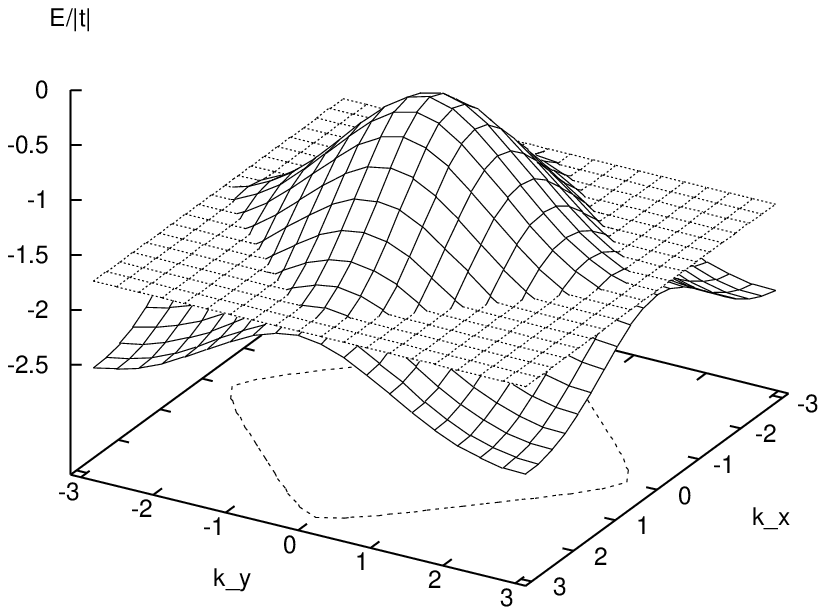}
\caption{The hole dispersion for $t>0$, $J/t=0.2$, $T=0.02t$,
$\eta=0.05t$ and $x=0.033$, 0.076, 0.14, and 0.18 (from top to bottom).
Horizontal planes are the Fermi levels. The respective Fermi surfaces
are shown in the base planes.}\label{Fig_i}
\end{figure}

\begin{figure}
\includegraphics[width=6.5cm]{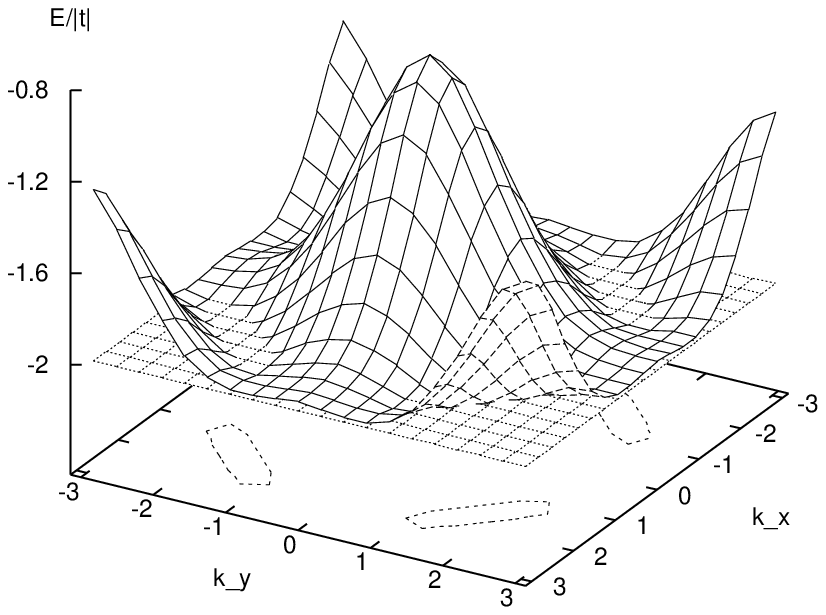}
\includegraphics[width=6.5cm]{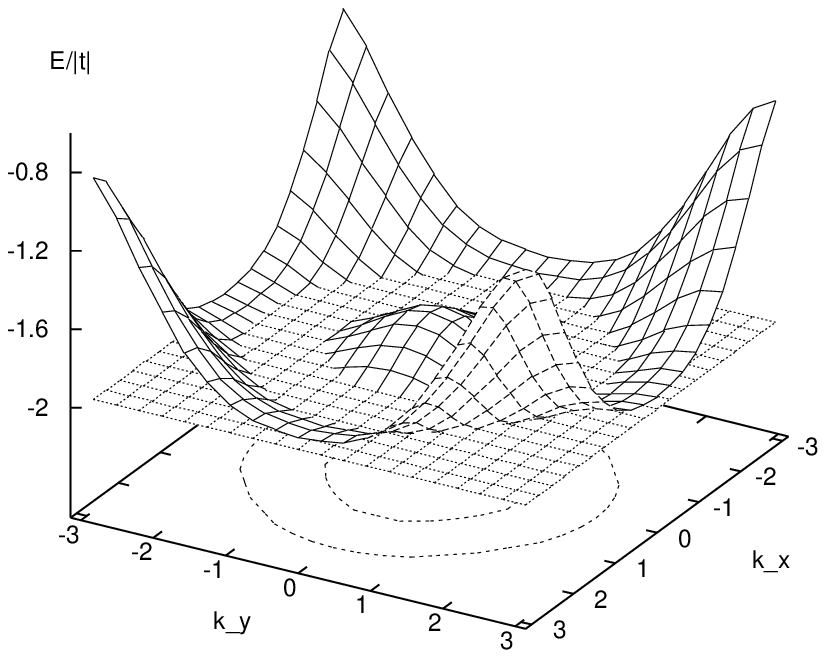}
\includegraphics[width=6.5cm]{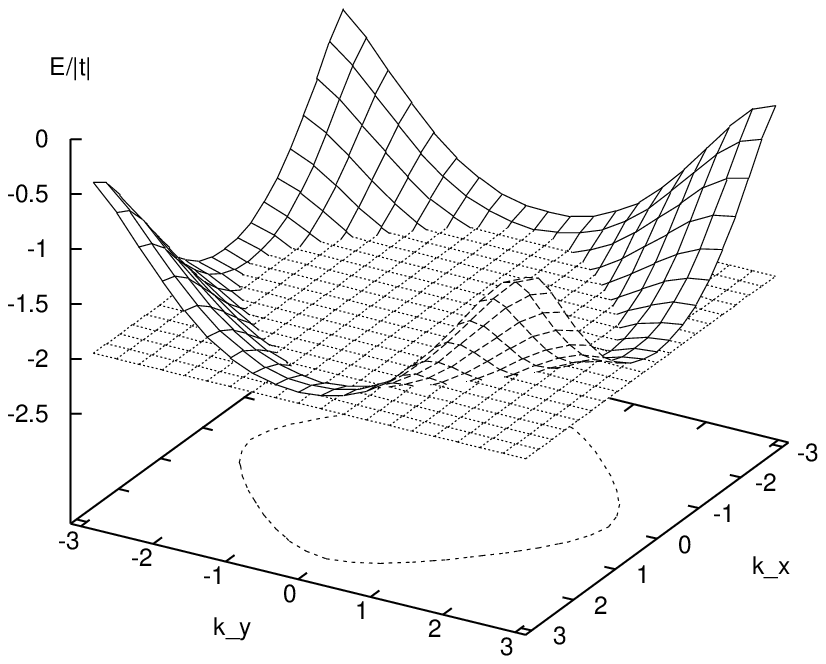}
\includegraphics[width=6.5cm]{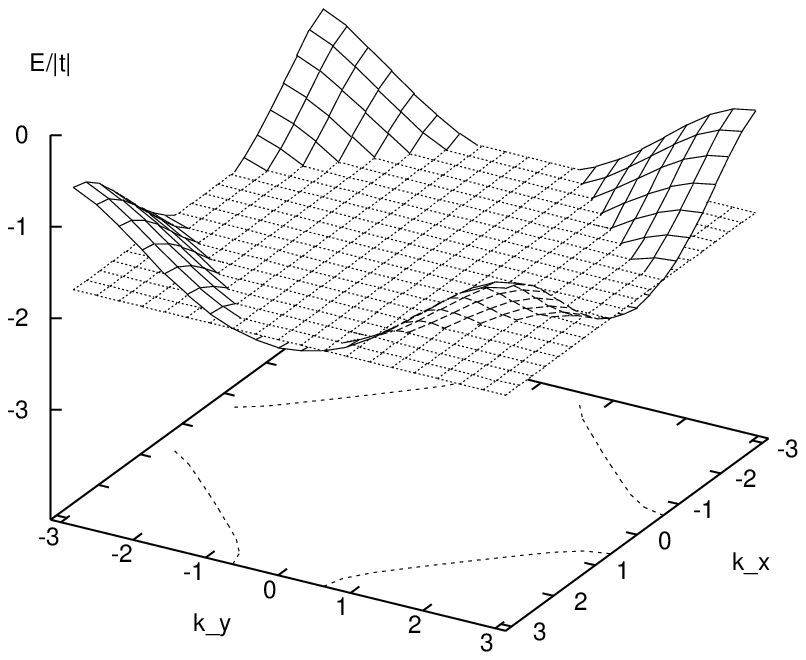}
\caption{The hole dispersion for $t<0$, $J/|t|=0.2$, $T=0.02|t|$,
$\eta=0.05|t|$ and $x=0.033$, 0.076, 0.13, and 0.2 (from top to
bottom).} \label{Fig_ii}
\end{figure}

Results of our calculations for the lowest of the hole bands which
crosses the Fermi level are shown in Figs.~\ref{Fig_i} and \ref{Fig_ii}
for different values of the hole concentrations and the sign of $t$.
Notice that for low hole concentrations the shape of the band is close
to the shape of the spin-polaron band obtained in the spin-wave
approximation \cite{Izyumov,Sherman98} and in the approach of
Ref.~\onlinecite{Sherman}. However, for the parameters chosen the
bandwidth is half as much again the value found in the two mentioned
approaches. This difference is a consequence of our two-pole
approximation (\ref{hgf}) for the hole Green's function. For small $x$
the shape of the band depends only weakly on the sign of $t$. With
increasing $x$ the band changes rapidly and already for moderate
concentrations its shapes for opposite signs of $t$ differ essentially
and resemble shapes of weakly correlated bands. As this takes place,
the bandwidth grows and for $x\approx 0.3$ it is three times as much as
the bandwidth for light doping.

The decrease of the temperature to $T=0.005|t|$, of the hole damping to
$\eta=0.015|t|$ or the increase of the exchange constant to $J=0.4|t|$
do not change qualitatively the evolution of the hole band with doping.
For all these parameters the transformation of the hole band from the
shape typical for strong correlations to that resembling the weakly
correlated case occurs at $x \approx 0.08$. As this takes place, the
respective Fermi surface, shown in the base planes in Figs.~\ref{Fig_i}
and \ref{Fig_ii}, changes its shape from small ellipses at
$(\pm\frac{\pi}{2},\pm\frac{\pi}{2})$ for small $x$ to a large rhombus
centered at $(\pi,\pi)$ or $(0,0)$ for large $x$. The van Hove
singularity of the hole band located at $(0,\pi)$ remains near the
Fermi level in the wide range of hole concentrations $x\alt 0.2$.

The reason for the above transformation of the band shape is in the
following. For small hole concentrations the hole dispersion is
determined by the magnetic ordering which is close to the long-range
antiferromagnetic order. In this case the elementary cell in the direct
space is doubled due to the opposite alignment of spins on neighboring
sites. Hence the Brillouin zone becomes half the value of the
paramagnetic zone and the points $(0,0)$ and $(\pi,\pi)$ become
equivalent. In particular, this means that the hole dispersion is
approximately invariant with respect to the substitution ${\bf k
\rightarrow k+Q}$, as seen in the upper parts of Fig.~\ref{Fig_i} and
\ref{Fig_ii}. Mathematically this corresponds to the case when
$|\widetilde{E}_0|\ll V_0$ in $\varepsilon_{{\bf k},j}$ in
Eq.~(\ref{energy}). With increasing $x$ the spin correlation length
$\xi$ is decreased. When \cite{Sherman} $\xi=\frac{1}{2}a\Delta^{-1/2}$
becomes comparable to a few lattice spacings $a$ the above arguments
cease to work. In this case the hole dispersion is determined by
$\widetilde{E}_0$ which becomes larger than $V_0$ and the band acquires
the shape which is similar to the case of weak correlations. However,
even for $x=0.25$ the bandwidth is much smaller than $8|t|$, the
bandwidth in the latter case (see the lower parts of Fig.~\ref{Fig_i}
and \ref{Fig_ii}). Thus, even in the overdoped case the corrections due
to electron correlations are essential. The above transformation is not
connected with some sharp transition in the magnetic subsystem. The
decrease of the spin correlation length with $x$ is continuous.

The analogous transformation of the hole band is expected also in the
more complicated and flexible $t$-$t'$-$t''$-$J$ model which is
frequently used for the interpretation of experimental results in
cuprates. \cite{Izyumov,Damascelli} The respective value of the hole
concentration is supposed to be close to $0.08$. This assumption is
based on the fact that the transformation occurs when the correlation
length approaches a few lattice spacings. The correlation length
derived from neutron scattering experiments \cite{Kastner} in cuprates
acquires this value near the mentioned concentration.

In the considered two-pole approximation only one hole band crosses the
Fermi level. In contrast to this the calculations based on the
spin-wave approximation \cite{Sherman98} and on the approach of
Ref.~\onlinecite{Sherman}, which are supposed to be more accurate for
low hole concentrations, show that for $x>0.05$ there are two bands
crossing the Fermi level. One of these bands is the spin-polaron band
which corresponds to a pronounced peak in the hole spectral function
and has the dispersion similar to that shown in upper parts of
Figs.~\ref{Fig_i} and \ref{Fig_ii}. The second band has a much weaker
and broader peak in $A({\bf k}\omega)$ and therefore its contribution
to the lower band of the two-pole approximation can be neglected. Thus,
the lower hole band which cross the Fermi level corresponds to the
feature with the highest peak intensity in the hole spectrum -- the
spin-polaron band. The upper band corresponds to ``everything else'' in
the hole spectrum. From previous considerations (see
Refs.~\onlinecite{Izyumov,Sherman98,Sherman} and references therein) it
is known that in $A({\bf k}\omega)$ for low $x$ several weaker and
broader maxima are located above the spin-polaron peak. These maxima
merge into one broad maximum at larger concentrations. The upper band
of the two-pole approximation gives a rough description for these
peculiarities of the hole spectrum. Since states near the Fermi level
are of main interest and this approximation gives a satisfactory
description for these states, it is used in the present work.

It is worth noting that the used two-pole approximation for the hole
Green's function gives satisfactory results not only in the limit of
small $x$ but also in the limit $x \rightarrow 1$. Indeed, in this
limit $V_0 \rightarrow 0$, $E_0 \approx 4t\gamma_{\bf k}-\mu$ and
$G({\bf k}\omega)=(\omega-4t\gamma_{\bf k}+\mu)^{-1}$. However, the
two-pole approximation becomes inapplicable in the intermediate region
for $x> 0.3$ when the chemical potential falls into the gap between two
bands.

\section{Magnetic susceptibility}
\begin{figure}
\includegraphics[width=6.5cm]{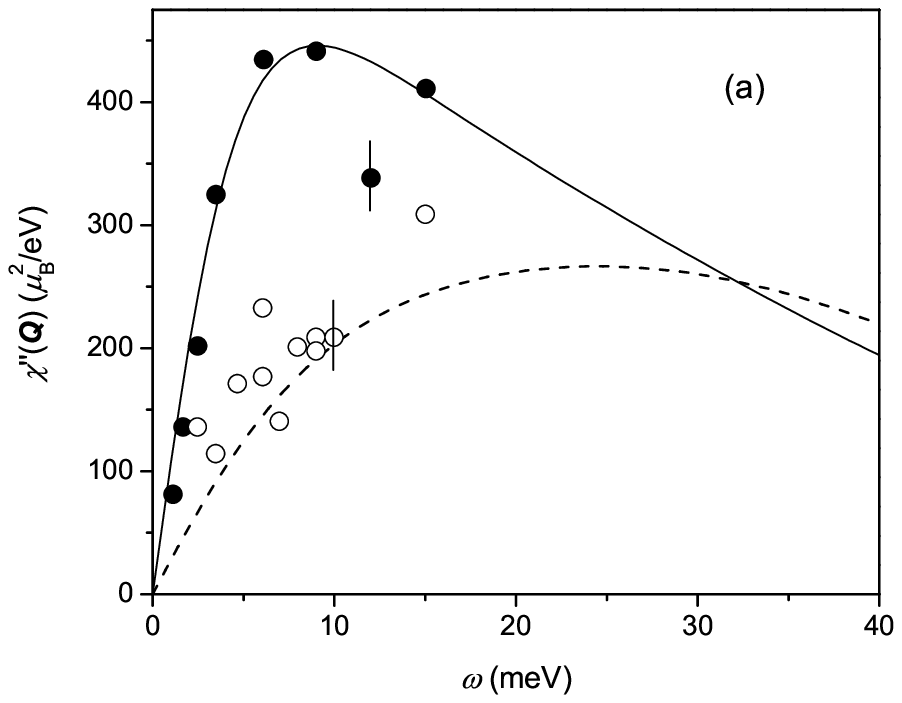}

\vspace{4ex}\includegraphics[width=6.5cm]{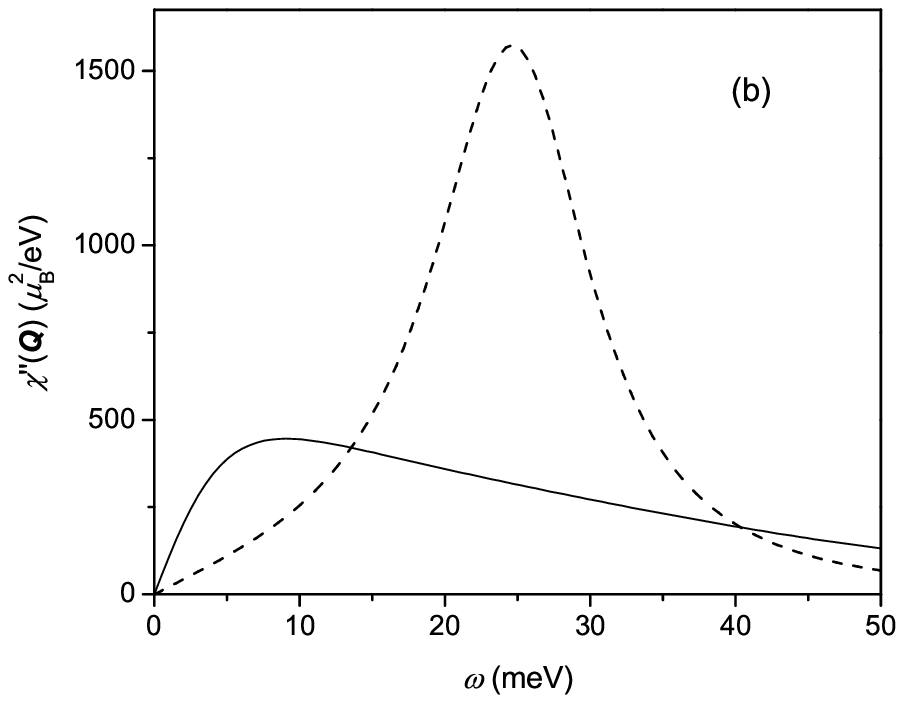} \caption{The
frequency dependence of the imaginary part of the susceptibility. (a)
The solid line corresponds to $J=0.2t$, $t>0$, $T=0.005t$, $\eta=0.05t$
and $x \approx 0.1$. The dashed line are for $T=0.02t$ and the same
other parameters. Symbols are experimental results
\protect\cite{Aeppli} in La$_{1.86}$Sr$_{0.14}$CuO$_4$ at $T=35$~K
(filled circles) and $T=80$~K (open circles). Vertical bars show
experimental errors. The calculated susceptibility is given for ${\bf
k}={\bf Q}$, the experimental data are for the wave vector of the
incommensurate peak. (b) The solid line is the same as in part (a). The
dashed line is calculated for $\eta=0.015t$, other parameters are the
same as for the solid line.} \label{Fig_iii}
\end{figure}

The imaginary part of the magnetic susceptibility is given by the
equation
\begin{eqnarray}
\chi''({\bf k}\omega)&=&4\pi\mu_B^2 B({\bf k}\omega)\nonumber\\
&=&\frac{16\mu_B^2(1-\gamma_{\bf k})(JC_1+tF_1)\omega{\rm
Im}\,\Pi}{(\omega^2-\omega{\rm Re}\,\Pi-\omega_{\bf k}^2)^2+(\omega{\rm
Im}\,\Pi)^2}, \label{chi}
\end{eqnarray}
where $\mu_B$ is the Bohr magneton. For the considered hole
concentrations the hole contribution to the susceptibility is much
smaller than the spin contribution. \cite{Sherman98} Up to $x \approx
0.13$ the momentum dependence of $\chi''$ is strongly peaked at {\bf Q}
(the use of a comparatively small 20$\times$20 lattice does not allow
us to describe the incommensurability of the magnetic response -- the
low-density inverse space and the frequency independent damping $\eta$
produce a minimum in the spin-excitation damping ${\rm Im}\,\Pi({\bf
k})$ at ${\bf k}={\bf Q}$ which is too shallow to give rise to
incommensurability \cite{Sherman,Sherman04}). The calculated frequency
dependencies of the susceptibility at this momentum are shown in
Fig.~\ref{Fig_iii}. In La$_{1.86}$Sr$_{0.14}$CuO$_4$ the susceptibility
is peaked at incommensurate momenta ${\bf k}=(\pi\pm 2\pi\delta,\pi),
(\pi,\pi\pm 2\pi\delta)$. The experimental susceptibility measured
\cite{Aeppli} at one of these momenta is also shown in
Fig.~\ref{Fig_iii}a. To demonstrate the similarity of the experimental
and calculated dependencies the former was increased by approximately
2.5 times (the difference in the peak amplitudes is apparently
connected with the splitting of the commensurate peak into 4
incommensurate maxima). As mentioned above, for the comparison with
experiment we set $t=0.5$~eV, thus the temperatures $T=0.005t$ and
$T=0.02t$ in the figure correspond to 29~K and 116~K, respectively.

As seen from Fig~\ref{Fig_iii}a, the calculated frequency dependencies
are close to those observed in La$_{1.86}$Sr$_{0.14}$CuO$_4$ for both
temperatures. For $T=0.005t$ the susceptibility has a broad maximum at
$\omega\approx 7$~meV. This maximum is not connected with the resonance
denominator in Eq.~(\ref{chi}), because for the parameters chosen spin
excitations near {\bf Q} are overdamped -- their dampings are larger
than their frequencies. A completely different situation occurs with
the decreased hole damping. The dampings of spin excitations decrease
also, the excitations cease to be overdamped and the shape of the
susceptibility is determined by the resonance denominator in
Eq.~(\ref{chi}). In this case $\chi''$ has a pronounced maximum at
$\omega \approx \omega_{\bf k}$, as seen in Fig.~\ref{Fig_iii}b. The
shape of the dashed curve in this figure is similar to the resonance
peak observed in the normal state of the underdoped
YBa$_2$Cu$_3$O$_{7-y}$. \cite{Bourges} Thus, we suppose that the
dissimilarity of the frequency dependencies of the susceptibility in
this crystal and in La$_{2-x}$Sr$_x$CuO$_4$ is connected with different
values of the damping of spin excitations. One of the possible reasons
for this difference is the diverse values of the hole damping.

\begin{figure}
\includegraphics[width=6.5cm]{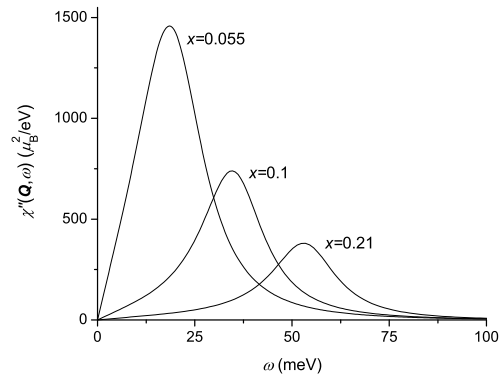}
\caption{The susceptibility at the antiferromagnetic wave vector for
the 3 hole concentrations indicated near the curves. Other parameters
are $J=0.2t$, $T=0.02t$, $\eta=0.02t$ and $t>0$.} \label{Fig_iv}
\end{figure}

The evolution of $\chi''({\bf Q}\omega)$ with the increase of the hole
concentration is shown in Fig.~\ref{Fig_iv}.
For the damping chosen all 3 curves have pronounced maxima, however
also for a larger damping, when spin excitations are overdamped, the
behavior of the susceptibility is similar to that shown in the figure.
With growing $x$ the frequency of the maximum increases and its
amplitude decreases. This behavior points to the increase of the spin
excitation frequency $\omega_{\bf Q}$ with the hole concentration. It
can be shown \cite{Kondo} that the value of this frequency is directly
connected with the magnetic correlation length $\xi$. From the obtained
results it follows that for low hole concentrations and temperatures
$\xi \approx ax^{-1/2}$ where $a$ is the intersite distance. An
analogous relation has been derived from experimental data in
La$_{2-x}$Sr$_x$CuO$_4$. \cite{Keimer} The peak amplitudes of the
susceptibility decreases rapidly with increasing $x$ for $x \alt 0.11$
and then flattens out. The rapid decrease can be related to the
transformation of the hole band discussed in the previous section. The
qualitative behavior of the peak amplitude is not changed with the
increase of the exchange constant to $J=0.4t$, with the decrease of the
temperature to $T=0.005t$ or with the change of the sign of $t$. The
qualitatively similar behavior of the susceptibility is observed
\cite{Bourges} in the normal-state YBa$_2$Cu$_3$O$_{7-y}$. However, in
experiment the decrease of the peak intensity is somewhat slower for
small $x$ and the maximum of the susceptibility is smeared out for
$x>0.16$. These features depend strongly on the damping of spin
excitations which is described only crudely in the used two-pole
approximation.

\begin{figure}
\includegraphics[width=6.5cm]{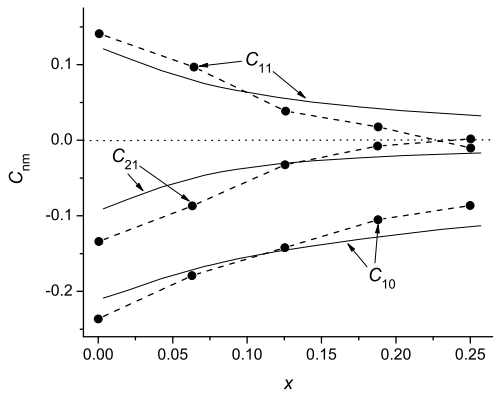}
\caption{The spin correlations $C_{\bf l}=\langle s^{+1}_{\bf
l}s^{-1}_{\bf 0}\rangle$, ${\bf l}=(n,m)$ obtained in our calculations
for $J=0.4t$, $T=0.005t$, $\eta=0.05t$ and $t>0$ (solid lines) and in
exact-diagonalization calculations \cite{Bonca} in a 4$\times$4 lattice
for $J=0.4t$ and $T=0$ (symbols, dashed lines are drawn as a guide for
the eye).} \label{Fig_v}
\end{figure}
The decrease of the susceptibility at the antiferromagnetic wave vector
with doping is reflected in the weakening of spin correlations $C_{\bf
l}=\langle s^{+1}_{\bf l}s^{-1}_{\bf 0}\rangle$. These correlations are
shown in Fig.~\ref{Fig_v} where the components of the vector ${\bf l}$
are designated as $n$ and $m$. In this figure the correlations obtained
\cite{Bonca} by exact diagonalization in a 4$\times$4 lattice for
$J=0.4t$ and $T=0$ are also shown. The difference between the two sets
of data is connected with the influence of finite-size effects in the
latter data and with the known issue of the method used in our
calculations which somewhat underestimates the correlations.
\cite{Sherman,Kondo} The correlations change only slightly with the
change of the sign of $t$ and with the variation of $J$ and $T$ in the
considered ranges. In spite of the large sensitivity of the spin
susceptibility on $\eta$, the spin correlations depend only weakly on
the hole damping due to the integration over frequencies in the formula
for the spin correlations (\ref{corr}). The change of the hole damping
from $\eta=0.05t$ to $0.015t$ leads to the variation in $C_{\bf l}$ by
several percent.

\section{Summary}
In this paper the evolution of the hole and spin-excitation spectra of
the two-dimensional $t$-$J$ model was studied in the range of hole
concentrations $0 \leq x \alt 0.3$ which spans the regions from light
to heavy doping. The variation of the spectra with temperature, with
the sign of the hopping parameter and with the excitation damping was
also considered. For this purpose the self-energy equations were
derived employing the projection operator technique and these equations
were self-consistently solved. The hole band was found to transform
radically at $x\approx 0.08$. A narrow low-concentration band with
minima near $(\pm\frac{\pi}{2},\pm\frac{\pi}{2})$ is converted to a
wider band resembling in shape the case of weak electron correlations,
with the minimum at $(\pi,\pi)$ or at $(0,0)$ in dependence on the sign
of $t$. The hole Fermi surface is respectively changed from small
elliptical pockets at $(\pm\frac{\pi}{2},\pm\frac{\pi}{2})$ to a large
rhombus centered at $(\pi,\pi)$ or $(0,0)$. The frequency dependence of
the imaginary part of the magnetic susceptibility $\chi''$ at the
antiferromagnetic wave vector depends heavily on the damping of spin
excitations and varies from a broad feature similar to that observed
\cite{Aeppli} in La$_{2-x}$Sr$_x$CuO$_4$ to a pronounced maximum which
resembles the normal-state resonance peak in YBa$_2$Cu$_3$O$_{7-y}$.
\cite{Bourges} One of the possible reasons for the variation of the
spin-excitation damping is the change of the hole damping. With
increasing doping the maximum in the susceptibility loses its intensity
and shifts to higher frequencies. The similar behavior is observed in
cuprates and is connected with the growth of the spin-excitation
frequency at the antiferromagnetic wave vector, which reflects the
decrease of the magnetic correlation length with doping.

\begin{acknowledgments}
This work was supported by the ESF grant No.~5548.
\end{acknowledgments}

\end{document}